\documentclass[pre,superscriptaddress,twocolumn,floatfix,nofootinbib]{revtex4}
\usepackage{amsmath}
\usepackage{amssymb}
\usepackage{graphicx}

\begin{document}

\title{A phenomenological theory giving
the full statistics of the position of fluctuating pulled fronts}

\author{E. Brunet}
\author{B. Derrida}
\affiliation{Laboratoire de Physique Statistique, \'Ecole Normale Sup\'erieure,
24 rue Lhomond, 75231 Paris cedex 05, France}
\author{A.~H. Mueller}
\affiliation{Department of Physics, Columbia University,
New York, NY 10027, USA}
\author{S. Munier}
\affiliation{Centre de Physique Th{\'e}orique, Unit\'e mixte de
recherche du CNRS (UMR 7644),
{\'E}cole Polytechnique, 91128~Palaiseau, France}
\affiliation{Dipartimento di Fisica, Universit\`a di Firenze, via Sansone 1,
50019~Sesto F., Florence, Italy.}

\date{\today}

\begin{abstract}
We propose a phenomenological description for the effect of a weak noise
on the position of a front described by the
Fisher-Kolmogorov-Petrovsky-Piscounov equation or any other travelling
wave equation in the same class. Our scenario is based on four hypotheses
on the relevant mechanism for the diffusion of the front. Our parameter-free
analytical predictions for the
velocity of the front, its diffusion constant and higher cumulants of its
position agree with numerical simulations.
\end{abstract}

\maketitle

\section{Introduction}

The Fisher Kolmogorov-Petrovsky-Piscounov (FKPP) equation \cite{FKPP}
\begin{equation}
\partial_t h = \partial_x^2 h + h - h^2,
\label{FKPP}
\end{equation}
describes how a stable phase ($h(x,t)=1$ for $x\to-\infty$) invades an
unstable phase ($h(x,t)=0$ for $x\to+\infty$) and how the front between
these two phases builds up and travels \cite{VanSaarloos.03}. This
equation was first introduced in a problem of genetics, but equations
similar to~(\ref{FKPP}) appear in much broader contexts like
reaction-diffusion
problems\cite{PechenikLevine.99,DoeringMuellerSmereka.03},
optimization\cite{MajumdarKrapivsky.03},
disordered systems\cite{DerridaSpohn.Polymers.88,CarpentierLedoussal.00}
and even particle
physics\cite{MunierPeschanski.03,BK,Marquet.05}.
A remarkable example is the problem of the high energy scattering of a
projectile consisting of a small color dipole on a target 
in the framework of quantum chromodynamics (QCD):
in Ref.~\cite{MunierPeschanski.03} it was recognized that the
Balitsky-Kovchegov (BK) equation \cite{BK}, 
a mean field equation for high energy 
scattering in QCD, is in the same class as the FKPP equation with $h$
being the scattering amplitude, $t$ the rapidity of the scattering and
$x$ the logarithm of the inverse projectile size. 

It is well known\cite{AronsonWeinberger.75,VanSaarloos.03} that equations like (\ref{FKPP}) have a family of
travelling wave solution of the form
$h(x,t)=h(z)$ with $z=x-vt$. 
There is a relation between the exponential decay of each solution
($h(z)\sim \exp(-\gamma z)$ for large $z$) and its velocity: $v=v(\gamma)$.
For example, $v(\gamma)=\gamma+1/\gamma$ for the
FKPP equation~(\ref{FKPP}). Other front equations would give
different expressions of $v(\gamma)$. See, for example,
section~\ref{sec:num}, or Refs.~\cite{BrunetDerrida.97,Enberg.05}.

If one starts with a steep enough initial condition, the front converges
to the travelling wave with the minimal velocity. Therefore
\begin{align}
v_\text{deterministic}&=\min_\gamma v(\gamma)=v(\gamma_0)\text{\quad
where\quad
}v'(\gamma_0)=0,\notag\\
h_\text{deterministic}(z) &\approx A z e^{-\gamma_0 z}.
\label{defgamma}
\end{align}
(The multiplicative factor $z$ in $h_\text{deterministic}$ is present
only for this slowest moving solution.)

There is a large class (the FKPP class) of equations describing the
propagation of a front into an unstable state which select the minimal
velocity, as described by (\ref{defgamma}). (There exist also equations,
called ``pushed'' or ``type~II'' for which the velocity selected by the
front is not the slowest one. The properties of these fronts are quite
different\cite{Kessler.98,VanSaarloos.03} from the properties of
(\ref{FKPP}), and we will not consider them in the present paper.)

Deterministic front equations such as~(\ref{FKPP}) usually occur as the
limit of a stochastic reaction-diffusion model \cite{Panja} when the
number of particles (or bacterias, or reactants) involved becomes
infinite. In a physical situation, all numbers remain finite and a small
noise term should be added to~(\ref{FKPP}) to represent the fluctuations
at the microscopic scale. One might write, for
instance\cite{MuellerSowers.95},
\begin{equation}
\partial_t h = \partial_x^2 h + h - h^2 + \sqrt{h(1-h)/N}\ \eta(x,t),
\label{noisy}
\end{equation}
where $\eta(x,t)$ is a normalized Gaussian white noise and $N$ is the
number of particles involved.

The effect of such a noise is to make the shape of the travelling wave
fluctuate in time\cite{DoeringMuellerSmereka.03}. It affects also its
velocity and makes the front diffuse\cite{BD2,Panja,VanSaarloos.03}.

For a chemical problem, $N$ might be of the order of the Avogadro number
and one could think that such a small noise term should give small
corrections, of order $1/\sqrt{N}$, to the shape and position of the
front. However, because the front motion is extremely sensitive to small
fluctuations in the region where $h\simeq1/N$, this is not the case. In
presence of noise as in (\ref{noisy}), the front has an exponential
decay if $h(x,t)\gg1/N$, but it vanishes much faster
than this exponential in the region where $h(x,t)$ is of order $1/N$~\cite{DoeringMuellerSmereka.03}. (This is obvious
in a particle model, as there cannot be less than one particle at a given
place.) As an approximation to understand the effect of the microscopic
stochastic details of the system, it has been suggested to replace the
noise term by a
deterministic cutoff which makes the front vanish very quickly when
$h\simeq1/N$ \cite{BrunetDerrida.97}. For instance, for the FKPP equation
(\ref{FKPP}), one way of introducing the cutoff is
\begin{equation}
\begin{gathered}
\partial_t h = \partial_x^2 h + \big(h - h^2\big)a(Nh),\\
\text{with $a(r)=1$ for $r>1$ and $a(r)\to0$ for $r\to0$}.
\end{gathered}
\label{FKPP-cutoff}
\end{equation}
In the presence of such a cutoff, the velocity and the shape
(\ref{defgamma}) become,
for any equation in the FKPP class,
\begin{subequations}
\begin{align}
h_\text{cutoff}(z)&\approx A { L\over\pi} \sin\left(\frac{\pi z}{L}\right)e^{-\gamma_0 z}
\text{\quad where\quad } L=\frac{1}{\gamma_0}{\log N},\label{h_cutoff}\\
v_\text{cutoff}&\approx v(\gamma_0)-{\frac{\pi^2 v''(\gamma_0)}{2 L^2}}.
\label{v_cutoff}
\end{align}
\label{cutoff}
\end{subequations}
(The shape (\ref{h_cutoff})
is valid only in the linear region, where $h$ is small
enough for the nonlinear term $h^2$ to be negligible but still larger
than $1/N$. Note that for $z\ll L$, the shape coincides with
(\ref{defgamma}). A way to interpret the sine is to say that
the front moves slower than the minimal
velocity~$v_\text{deterministic}=v(\gamma_0)$ and that the decay rate
becomes complex: $\gamma=\gamma_0 \pm
i\pi/L$. Then, the expression of $v_\text{cutoff}$ results from an expansion of
$v(\gamma)$ for large $L$.)

The prediction (\ref{cutoff}) does not depend on the details of the
microscopic model. It only depends on the deterministic equation and on
the existence of a microscopic scale. This cutoff picture is also present
in the mean field QCD context in~\cite{MuellerShoshi.04}, where it was
introduced to avoid unitarity violating effects in the BK equation at
intermediate stages of rapidity evolution. In this context, $N$ is
$1/\alpha_\text{QCD}^2$ where $\alpha_\text{QCD}$ is the strong coupling
constant.

Extensive numerical simulations of noisy fronts have been
performed over the years\cite{BD2,PechenikLevine.99}, and the large
correction~(\ref{v_cutoff}) to the velocity found in the cutoff picture seems to give the
correct leading correction to the velocity of noisy fronts. (See
\cite{ColonDoering.05} for rigorous bounds.)
Being a deterministic approximation, the cutoff theory gives however
no prediction for the diffusion constant of the front.

In the present paper, we develop a phenomenological description which leads to a
prediction for this diffusion constant. This description tries to capture
the rare relevant events which give the dominant contribution to the
fluctuations in the position of the front. The prediction is that
the full statistics of the front position in the noisy model depends only
on the amplitude~$1/N$ of the noise at the microscopic scale and on
$v(\gamma)$, a
property of the deterministic equation. For large~$N$,
all the other details of the
underlying microscopic model do not contribute to the leading order.
Our description leads to the following
prediction for the velocity and for the diffusion constant of
the front for large $N$: 
\begin{subequations}\label{results}
\begin{equation}\boxed{\begin{aligned}
v - v_\text{cutoff} &={\pi^2\gamma_0^3 v''(\gamma_0)}\
  \frac{3\log\log N}{\gamma_0 \log^3N}+\cdots\\
D &= {\pi^2\gamma_0^3 v''(\gamma_0)}\ {\frac{\pi^2/3}{\gamma_0^2\log^3N}}
+\cdots
\label{resultsvD}
\end{aligned}}\end{equation}

Actually, our phenomenological approach also gives a prediction to the
leading order for all the cumulants of the position of the front.
For $n\ge2$,
\begin{equation}\boxed{
\frac{[\text{$n$-th cumulant}]}{ t}
 = {\pi^2 \gamma_0^{3} v''(\gamma_0)}\ 
\frac{n!\zeta(n)}{\gamma_0^n\log^3N}+\cdots
}
\label{resultscumulants}
\end{equation}
\end{subequations}
where $\zeta(n)=\sum_{k\ge1} k^{-n}$.

The $1/\log^3N$ dependence of the diffusion constant was already observed
in numerical simulations \cite{BD2}. In the QCD context, it was proposed in
\cite{IancuMuellerMunier.05} to identify the full QCD problem with a stochastic evolution,
such as (\ref{noisy}), and the dependence of the diffusion constant was
used to suggest a new scaling law for QCD hard scattering at, perhaps,
ultrahigh energies.

We do not have, at present, a mathematical proof of the results
(\ref{resultsvD}) and
(\ref{resultscumulants}). Rather, we believe
that we have identified the main effects contributing to the diffusion of
the front.
We present our scenario in Sec.~\ref{sec:deriv} where we state a set of
four hypotheses from which the results~(\ref{results}) follow.
We give arguments to support these hypotheses
in sections~\ref{sec:j1}
to~\ref{sec:j4}. Finally, to check our claims,  we present numerical
simulations in section~\ref{sec:num} for the five first cumulants of the
position of the front. These simulations match very well the
predictions~(\ref{results}).

\section{The picture and its quantitative consequences}
\label{sec:deriv}
To simplify the discussion, we consider, in this section, more
specifically a microscopic particle model rather than a continuous
stochastic model such as (\ref{noisy}). This is merely a convenience to
make our point clearer, but the discussion below
could be rephrased for other models in the stochastic FKPP class.

We consider models where particles diffuse on the line and,
occasionally, duplicate. If one considers, for $h(x,t)$,
the density of particles or,
alternatively, the number of particles on the right of $x$, it is clear
that it is not yet described by a front equation, because it grows
exponentially fast with time; one needs to introduce a saturation rule. For
instance: 1) keep the number of
particles fixed by removing the leftmost particles if necessary; or 2)
remove all the particles which are at a distance larger than~$L$ behind
the rightmost particle; or 3) limit the density by allowing,
with a small probability, that two particles meeting recombine into
one single particle\cite{DoeringMuellerSmereka.03}.

\subsection{A scenario for the propagation of the front}
The main picture of our phenomenological description
is the following. The evolution of the front is
essentially deterministic, and its typical shape and velocity are given
by Eq.~(\ref{cutoff}). But from time to time, a fluctuation sends a
small number of particles at some distance $\delta$ ahead of the front.
At first, the position of the front, determined by where most of the
particles are, is only modified by a negligible amount of order $1/N$ by
this fluctuation. However, as the system relaxes, the number of wandering
particles grows exponentially and they start contributing to the position
of the front. Meanwhile the bulk catches up and absorbs the wandering
particles and their many offsprings; finally, the front relaxes back to
its typical shape (\ref{h_cutoff}). The net effect of a fluctuation is
therefore to shift the position of the front by some amount $R(\delta)$
which depends, obviously, on the size $\delta$ of the fluctuation. A
useful quantity to characterize the fluctuations is the width of the
front. It can easily be defined as the distance between the leading
particle (where $h \approx 1/N$) and some position in the bulk of the
front, for instance, where $h = 0.5$. (Changing this reference point would
change the width by a finite amount, independent of $N$). This width is
typically of order $L$, where $L$ is given by the cutoff theory
(\ref{h_cutoff}). During a fluctuation that sends particles at a distance
$\delta$ ahead of the front, the width of the front increases quickly to
$L+\delta$, and then relaxes slowly back to $L$.

We emphasize that, in this scenario, the effect of noise is so
weak that, most of time, it can be ignored and the cutoff theory
describes accurately the evolution of the front. It is only occasionally,
when a rare sequence of random microscopic events sends some particles
well ahead of the front that the cutoff theory is no longer valid. The
way this fluctuation relaxes is, however, well described by the
deterministic cutoff theory.

We shall encode this scenario in the following quantitative assumptions:
\begin{enumerate}
\item 
If we write the instantaneous fluctuating width of the front as
$L+\delta$, then the probability distribution function for $\delta$ is
given by
\begin{equation}
p(\delta)\,d\delta = C_1 e^{-\gamma_0\delta}\,d\delta,
\label{distlen}
\end{equation}
where $C_1$ is some constant.
Note that we assume this form only over some relevant range of values:
$\delta$ large enough (compared to 1) but much smaller than $L$
(typically of order $\log L$). Fluctuations where $\delta$ is ``too
small'' are frequent but do not contribute much to the front position.
Fluctuations where $\delta$ is ``too large'' are so rare that we do not
need to take them into account. Only for ``moderate'' values of $\delta$
do we assume the above exponential probability distribution function.

\item The long term effect of a fluctuation of size~$\delta$ (assuming
that there are no other fluctuations in-between) is a shift of the front
position by the quantity
\begin{equation}
R(\delta)=\frac{1}{\gamma_0}\log\left(1+C_2 
\frac{e^{\gamma_0 \delta}}{ L^3}
\right),
\label{Rd}
\end{equation}
where $C_2$ is another constant.

\item The fluctuations of the position of the front are dominated by
large and rare fluctuations of the shape of the front. We assume that
they are rare enough that a given relevant fluctuation has enough time to
relax before another one occurs.
\end{enumerate}

From these three hypotheses alone, one can
derive our results~(\ref{results}) up
to a single multiplicative constant. This constant can be determined with
the help of a fourth hypothesis:
\begin{enumerate}\setcounter{enumi}{3}
\item For the aim of computing the first correction to the front velocity
obtained in the cutoff theory (\ref{cutoff}), one can simply 
use the expression (\ref{v_cutoff}) with $L$ replaced
by $L_\text{eff}$ where
\begin{equation}
L_\text{eff}= \frac{1}{\gamma_0}\log N+\frac{3}{\gamma_0}\log\log N+\cdots
\label{4th}
\end{equation}
It is important to appreciate that the average or typical width of the front is
still $L$ and not $L_\text{eff}$. The latter quantity is just what
should be used in (\ref{v_cutoff}) to give the correct velocity.

\end{enumerate}

\subsection{How (\ref{results}) follows from these hypotheses}

We are now going to see how the results (\ref{results}) follow from these
four hypotheses.

First, we argue that the probability to observe a fluctuation of size
$\delta$ during a time interval $\Delta t$ can be written as
$p(\delta)d\delta\ \Delta t/\tau$, where $p(\delta)$ is the
distribution~(\ref{distlen})
of the increase of the width of the front and where $\tau$ is some
typical time characterizing the rate at which these fluctuations
occur. Indeed, during a fluctuation of a given size, the width of the
front increases to that size and then relaxes back. For a
large $\delta$, observing a front of size $L+\delta$ is very rare, but,
when it happens, the most probable is that one is observing the maximum
expansion of a fluctuation with a size close to $\delta$; the
contribution from fluctuations of sizes significantly larger than
$\delta$ is negligible as they are much less likely.

Second, as a fluctuation builds up at the very tip of the
front where the saturation rule (see beginning of
section~\ref{sec:deriv}) can be neglected, we argue that the typical time
$\tau$ introduced in the previous paragraph 
and the time it takes to build a fluctuation of a given size do
not depend on~$N$. (However, the relaxation time of a
fluctuation depends on~$N$ as
the bulk of the front is involved in the relaxation.)

Let~$X_t$ be the position of the front, $\delta_0$ the minimal size of
a fluctuation giving a relevant contribution to the position of the front
and~$\Delta t$ a time much
smaller than the time between two relevant fluctuations, but much larger
than the time it takes to build up such a fluctuation and have it relax.
(This is authorized by the third hypothesis.) We have
\begin{multline*}
X_{t+\Delta t}=\\\begin{cases}
X_t + v_\text{cutoff}\Delta t +R(\delta) &\text{proba. 
$\frac{\Delta t}{\tau}p(\delta)d\delta$}\text{ for
$\delta>\delta_0$},\\[1ex]
X_t + v_\text{cutoff}\Delta t &\text{proba. 
$1-\frac{\Delta t}{\tau}\int_{\delta_0}^\infty p(\delta)d\delta$}.
\end{cases}
\end{multline*}
(Note that ${\Delta t\over\tau}\int_{\delta_0}^\infty p(\delta)d\delta$ is the
probability of observing a relevant fluctuation during the time $\Delta
t$. By definition of $\Delta t$, this is much smaller than 1.)

One can then compute the average,  denoted $\langle\cdot\rangle$,
of $\exp(\lambda X_{t+\Delta t})$. One gets, for $\lambda$ small enough,
\begin{equation}
\partial_t \log\left\langle e^{\lambda X_t}\right\rangle
=\lambda v_\text{cutoff}+\frac{1}{\tau}\int p(\delta)\left[e^{\lambda
R(\delta)}-1\right]\ d\delta.
\end{equation}
Expanding in powers of~$\lambda$, one recognizes on the left-hand-side
the cumulants of~$X_t$. Therefore,  one gets
\begin{equation}
\begin{aligned}
v-v_\text{cutoff}&=\frac{1}{\tau}\int p(\delta)R(\delta)\ d\delta,\\
\frac{[\text{$n$-th cumulant}]}{t}
	&=\frac{1}{\tau}\int p(\delta)R^n(\delta)\
d\delta\quad\text{for $n\ge2$}.
\end{aligned}\label{integrals}\end{equation}
At this point, one can notice from the expressions of~$p(\delta)$ and
$R(\delta)$ that the values of $\delta$ such that
$e^{\gamma_0\delta}\gg L^3$ have a negligible contribution to the
integrals giving the velocity and the cumulants. Thus appears naturally a
$\delta_\text{max}=(3/\gamma_0)\log L$ which is exactly the
effective correction to the width of the front appearing in (\ref{4th}).

The integrals~in Eq.~(\ref{integrals}) can be evaluated, and one gets
\begin{equation}
\int p(\delta)R^n(\delta)\ d\delta=\frac{C_1 C_2}{\gamma_0^{n+1} L^3}
\int_0^\frac{L^3}{C_2}\log^n\left(1+\frac{1}{x}\right)\ dx,
\end{equation}
with~$x=(L^3/C_2)\exp(-\gamma_0\delta)$. For~$n=1$, this integral
gives~$\log(L^3/C_2)$. For~$n\ge2$, one can integrate from 0 to~$\infty$
(the correction is at most of order~$1/L^6$) and one
recognizes~$n!\zeta(n)$. Finally,
\begin{equation}
\begin{aligned}
v-v_\text{cutoff}&=\frac{C_1 C_2}{\tau\gamma_0}\ 
\frac{3\log L}{\gamma_0 L^3},\\
\frac{[\text{$n$-th cumulant}]}{t}
	&=\frac{C_1 C_2}{\tau\gamma_0}\ \frac{n!\zeta(n)}{\gamma_0^n L^3}.
\end{aligned}\end{equation}
Everything is determined up to \emph{one} numerical
constant~$C_1C_2/\tau$.
As the fourth hypothesis gives the velocity, one can easily determine that
constant and recover~(\ref{results}).

All the cumulants (except the first one) are of the same order of
magnitude, as the fluctuations are due to rare big events.

\section{Arguments to support the hypotheses}\label{sec:justification}%

\subsection{First hypothesis}\label{sec:j1}

This first hypothesis is not very surprising if one considers that
$\exp(-\gamma_0\delta)$ is the natural decay rate of the deterministic
equation.
A more quantitative way to understand~(\ref{distlen}) is that building up
a fluctuation is an effect which is very localized at the tip of the
front, where saturation effects can be neglected. We present in
appendix~\ref{infinite} a calculation using this property.

Moreover, numerical simulations\cite{Moro_private} of that probability
distribution function give evidence that for large enough $N$, the
decay is exponential with the rate~$\gamma_0$ as in~(\ref{distlen}).

\subsection{Second hypothesis}\label{sec:j2}

To obtain (\ref{Rd}), we need to compute the response of the
deterministic model with a cutoff (\ref{FKPP-cutoff}) to a fluctuation at
the tip of the front. This is a purely deterministic problem:
starting with a fluctuation (\emph{i.e.\@} a configuration slightly
different from the stationary shape), we let the system evolve with a
cutoff and relax back to its stationary shape (\ref{h_cutoff}), and we
would like to compute the shift in position due to this fluctuation.

Although the evolution is purely deterministic, the problem remains a
difficult one. For simplicity, we discuss here the case of the
FKPP equation (\ref{FKPP-cutoff}). The extension to other travelling wave
equations in the FKPP class is straightforward.

There are two non-linearities in (\ref{FKPP-cutoff}): one is the $-h^2$
term, which is important when $h$ is of order 1, and the other one is
the cutoff term $a(Nh)$, which is important when $h$ is of order~$1/N$.
Between these two points, there is a large length of order
$L=\log N$ where one can
neglect both non-linearities. This means that, for all practical purpose,
one  can simply use the linearized version of the FKPP equation for the
whole front except for two small regions with a size of order 1 at both
ends of the front.

Let $X_t$ be the position of the front, and $L_t$ its length. There are
many equivalent ways of defining precisely these quantities; for instance
we can take $X_t$ such that $h(X_t,t)=10^{-5}$ and $L_t$ such that
$h(X_t+L_t,t)={1\over N}$. We expect that $X_t-v_\text{cutoff}\,t$ and
$L_t-L$, which are quantities of order 1, have a relaxation time of
order~$L^2$, as for the shape of the front.

For $X_t<x<X_t+L_t$, the problem is linear:
\begin{equation}
\partial_t h = \partial_x^2 h + h.
\end{equation}
Using the Ansatz
\begin{equation}
h(x,t)= L_t G\bigg(\underbrace{x-X_t\over L_t}_y,\underbrace{t\over
L^2}_\tau\bigg)e^{-(x-v_\text{cutoff} t)},
	\label{ansatz}
	\end{equation}
	with $v_\text{cutoff}=2-{\pi^2\over L^2}$ (see (\ref{v_cutoff})
for $v(\gamma)=\gamma+1/\gamma$),
and keeping only the dominant terms in~$L$, the function~$G(y,\tau)$
evolves according to
\begin{equation}
\partial_\tau G=\partial_y^2 G +\pi^2 G,
\end{equation}
with the boundary conditions
\begin{equation}
G(0,\tau)\approx0,\quad	G(1,\tau)\approx0.
\end{equation}
(More precisely, $G(0,\tau)$ and $G(1,\tau)$ would be non zero only at the
next order in a $1/L$ expansion.)

The problem reduces to a diffusion problem with absorbing boundary
conditions. The stationary configuration is the sine shape (\ref{h_cutoff}),
as expected.

If, at time $t=0$ the shape is different from this stationary
configuration, it will relax back to it in the long time limit 
up to a multiplicative constant:
\begin{equation}
G(y,\infty) = {B\over\pi} \sin (\pi y).
\end{equation}
As the stationary shape for $h(x,t)$ must be of the form given by
(\ref{h_cutoff}), we obtain, using (\ref{ansatz}) that the final shift in
position is given by
\begin{equation}
R(\delta)=\lim_{t\to\infty} \big( X_t - v_\text{cutoff}\,t\big) = 
\log {B\over A}.
\end{equation}

To compute the value of $B$, one simply needs to project the initial
condition on the sine shape:
\begin{equation}
B=A e^{R(\delta)}=2\pi\int_0^1 dy\ \sin(\pi y) G(y,0).
\label{firstmode}
\end{equation}

We now proceed to use this expression for the perturbations we are
interested in: perturbations localized near the cutoff. 

We do not have a full information on the initial condition $h(x,0)$ or,
equivalently, $G(y,0)$. However, as we expect a perturbation to grow at
the very tip of the front, we expect that $h(x,0)$ is identical to its
stationary shape, except in a region of size of order $\Delta x\approx1$
on its tip. On the scale we consider, this means that $G(y,0)$ is
perturbed over a region of size $\Delta y\approx 1/L$. In other words:
\begin{equation}
G(y,0) = A\left[{1\over\pi}\sin(\pi y) + p(1-y)\right],
\end{equation}
where the perturbation $p(y')$ is non-zero only for $y'=1-y$ of
order~$1/L$. Therefore, from (\ref{firstmode}),
\begin{equation}
e^{R(\delta)}=1+2\pi\int_0^{b\over L} dy'\ \pi y' p(y'),
\end{equation}
where $b$ is a number of order 1 representing the extent over which a
perturbation initially affects the shape of the front. 
($p(y')\approx0$ if $y'>b/L$.)

The precise shape of $p(y')$ is not known, but its amplitude
can be easily understood
in a stochastic particle model: if some particles are sent at a distance
$\delta\ll L$ ahead of the front, $h(x,t)$ increases by $1/N$ at position
$x=X_t+L+\delta$. Because of the exponential factor in (\ref{ansatz}),
this translates to an increase of order $p(y')\approx\exp(\delta)/L$ for the
reduced shape $G(y,\tau)$.
Combining everything, one finally gets
\begin{equation}
e^{R(\delta)}=1+C_2 {e^\delta\over L^3},
\label{eRdelta}
\end{equation}
where $C_2$ is some number of order~1 which depends on the precise shape
$p(y')$. Expression~(\ref{eRdelta}) is just our second hypothesis,
up to factors~$\gamma_0$ which can be put back by dimensional
analysis.

One consequence of the argument above is
that $C_2$ is of order~1 compared to $L$. However, it gives no
information about the dependence of $C_2$ on $\delta$ or on the shape of
the fluctuation. We think that if $C_2$ depends on $\delta$, it is a weak
dependence that we can ignore. A simple situation where this can be
checked is when $\delta$ is large: if a particle jumps sufficiently far
ahead, it will start a front of its own that will completely replace the
original front. For such a front, it is well
known\cite{Bramson.83,BrunetDerrida.97} that the position for large~$t$
is given at first (while the cutoff is not relevant) by
$\delta+ 2 t-\frac{3}{2}\log t$.
When the velocity $2-\frac{3}{2t}$ matches $v_\text{cutoff}$, that is at a
time $t_0\approx L^2$, a crossover occurs and the position becomes
$R(\delta)+v_\text{cutoff}\,t$. Matching the two expressions for the
position at time $t=t_0$, one obtains
$R(\delta)\approx \delta-\log L^3$, as predicted by~(\ref{Rd}).
This indicates that, at least for large $\delta$, the number
$C_2$ has no $\delta$
dependence.

\subsection{Third hypothesis}\label{sec:j3}

From section~\ref{sec:deriv} and Eq.~(\ref{Rd}), the size~$\delta$ of the
fluctuations that contribute significantly to the diffusion of the front
are such that~$\exp(\gamma_0\delta)\sim L^3$. From (\ref{distlen}), the
typical time between two such fluctuations is therefore~$L^3$. On the
other hand, from section~\ref{sec:j2}, the relaxation time of
a fluctuation is of order~$L^2$. It is therefore safe to assume that a
relevant fluctuation has enough time to relax before another one
occurs.

\subsection{Fourth hypothesis}\label{sec:j4}

The fourth hypothesis states that, to compute the shift in velocity, one
should use a front width $L_\text{eff}$ that is larger than what is
predicted by the cutoff theory by an amount $\frac{3}{\gamma_0}{\log\log
N}$. The hypothesis is plausible as this length is precisely the
distance~$\delta$
at which the relevant fluctuations occur:
the main effect of the fluctuations would then be to increase
the effective width of the front that enters the cutoff
theory~(\ref{cutoff}). We present in appendix~\ref{wall} a simplified
model to support this claim.

Remarkably, the front width $L_\text{eff}$
emerges naturally in the QCD context\cite{MuellerShoshi.04}. 

\section{Numerical simulations}\label{sec:num}%

We consider here a reaction-diffusion model with saturation which was
introduced in~\cite{Enberg.05} as a toy model for high energy scattering
in QCD. Particles are evolving in discrete time 
on a one-dimensional lattice. At each timestep, a particle may jump
to the nearest position on the left or on the right with respective
probabilities $p_l$ and $p_r$, and may divide into two particles with
probability $\lambda$. We also impose that each of the $n(x,t)$ particles
piled up at $x$ at time $t$ may die with probability
$\lambda n(x,t)/N$.

\begin{figure*}[ht]
\centering
\includegraphics[width=.9\textwidth]{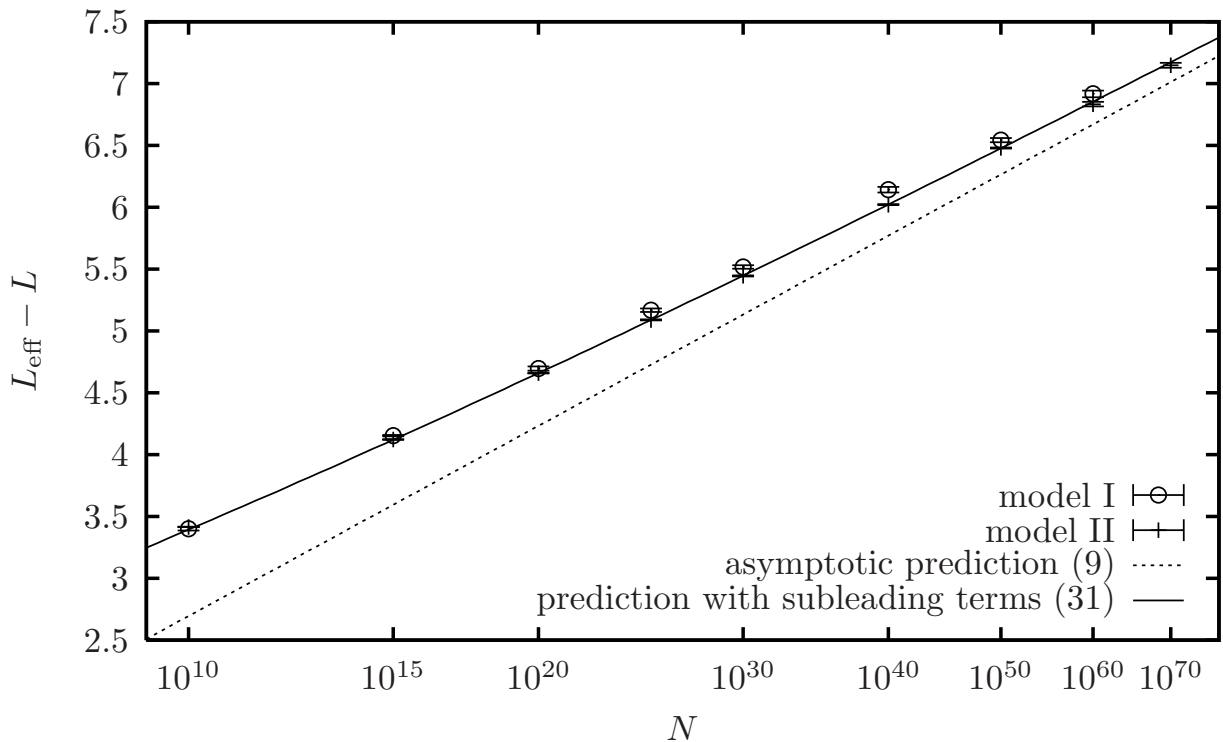}
\caption{\label{loglogN}
Measured $L_\text{eff}$, defined by~(\ref{Leff_from_v}),
from which we have subtracted 
the width $L$ in the cutoff theory, as a function of $N$.
The dashed line represents the leading terms $3\log\log N/\gamma_0$, see
(\ref{4th}).
The subleading terms (\ref{ansatzLeff})
of the plain line have been determined by a fit.}
\end{figure*}

Between times $t$ and $t+1$, $n_l(x,t)$ particles 
out of $n(x,t)$ move to the left and
$n_r(x,t)$ move to the right. 
Furthermore, $n_+(x,t)$ particles are replaced by their
two offsprings at $x$, and $n_-(x,t)$ particles disappear.
Hence the total variation in the number of particles on site $x$ reads
\begin{subequations}\label{stocha}
\begin{multline}
n(x,t+1)-n(x,t)
=-n_l(x,t)-n_r(x,t)-n_-(x,t)\\
+n_+(x,t)+n_l(x+1,t)+n_r(x-1,t).
\end{multline}
The numbers describing a timestep at position $x$ have 
a multinomial distribution:
\begin{multline}
P(\{n_l,n_r,n_+,n_-\})=
\frac{n!}{n_l!n_r!n_+!n_-!\Delta n!}
p_l^{n_l}p_r^{n_r}\\
\lambda^{n_+}(\lambda n/N)^{n_-}
(1\!-\!p_l\!-\!p_r\!-\!\lambda \!-\!\lambda n/N)^{\Delta n},
\end{multline}
\end{subequations}
where $\Delta n=n-n_l-n_r-n_+-n_-$, and all quantities 
in the previous equation
are understood at site $x$ and time $t$.
The mean evolution of $u\equiv n/N$ in one step of time reads
\begin{multline}
\langle u(x,t\!+\!1)|\{u(x,t)\}\rangle\!=\!u(x,t)\!+\!
p_l[u(x\!+\!1,t)\!-\!u(x,t)]\\
+\!p_r[u(x\!-\!1,t)\!-\!u(x,t)]\!+\!\lambda u(x,t)[1\!-\!u(x,t)].
\label{mod_mf}
\end{multline}
When $N$ is infinitely large, one can replace the $u$'s in~(\ref{mod_mf})
by their averages. One obtains then a deterministic front equation in the
FKPP class with
\begin{equation}
v(\gamma)=\frac{1}{\gamma}\log\left(1+\lambda+p_l(e^{-\gamma}-1)+p_r(e^{\gamma}-1)\right),
\label{vg}
\end{equation}
and $\gamma_0$ is defined by $v^\prime(\gamma_0)=0$, see~(\ref{defgamma}).

For the purpose of our numerical study, we set
\begin{equation}
p_l=p_r=0.1\text{\quad and\quad}\lambda=0.2\,.
\end{equation}
From (\ref{vg}), this choice leads to
\begin{equation}
\begin{split}
&\gamma_0=1.3521\cdots\ ,\ \ v(\gamma_0)=0.25538\cdots,\\
&v^{\prime\prime}(\gamma_0)=0.16773\cdots.
\end{split}
\end{equation}
Predictions for all cumulants 
of the position of the front 
are obtained
by replacing the values of these parameters in~(\ref{results}).

\begin{figure*}[ht]
\centering
\includegraphics[width=.9\textwidth]{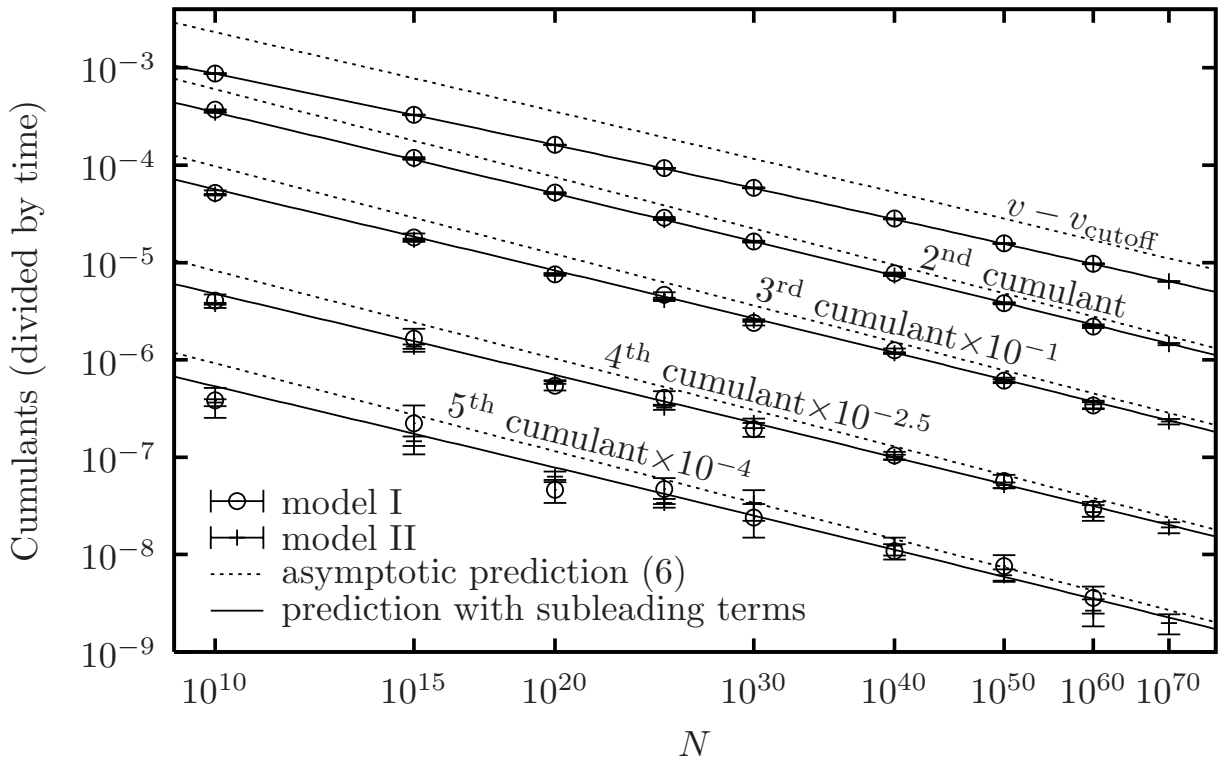}
\caption{\label{cumulants}From top to bottom, the correction to the
velocity given by the cutoff theory 
and the cumulants of orders 2 to 5 of the position
of the front in the stochastic model. The numerical data
are compared to our parameter-free 
analytical predictions~(\ref{results}), represented by the dashed line.
The subleading terms of the plain
lines are numerically the same as in figure~\ref{loglogN}; no further fit
has been performed for the present figure.}
\end{figure*}

Technically,
in order to be able to go to very large values of $N$, we replace the 
full stochastic model by its deterministic
mean field approximation $u\rightarrow \langle u\rangle$, where $\langle u\rangle$
is given by Eq.~(\ref{mod_mf}), in all bins in which the number of
particles is larger than $10^3$ (that is, in the bulk of the front).
Whenever the number of particles is smaller, we use the full 
stochastic evolution~(\ref{stocha}).
We add an appropriate boundary condition 
on the interface between the
bins described by the deterministic equation and the bins described by
the stochastic equation so that the flux of particles is
conserved\cite{Moro}.
This setup will be called ``model~I''.
Eventually, we shall use the 
mean field approximation
everywhere except in the rightmost bin (model~II): 
at each time step, a new bin is filled immediately
on the right of the rightmost nonempty site with a number 
of particles given by a Poisson law
of average
$\theta=N\langle u(x,t\!+\!1)|\{u(x,t)\}\rangle.$
In the context of
a slightly different model in the same universality class~\cite{BD2},
this last approximation was shown numerically to give indistinguishable results from
those obtained with the full stochastic version of the model, as far
as the front velocity and its diffusion constant were concerned.
We shall confirm this observation here.

\begin{figure*}[ht]
\centering
\includegraphics[width=.9\textwidth]{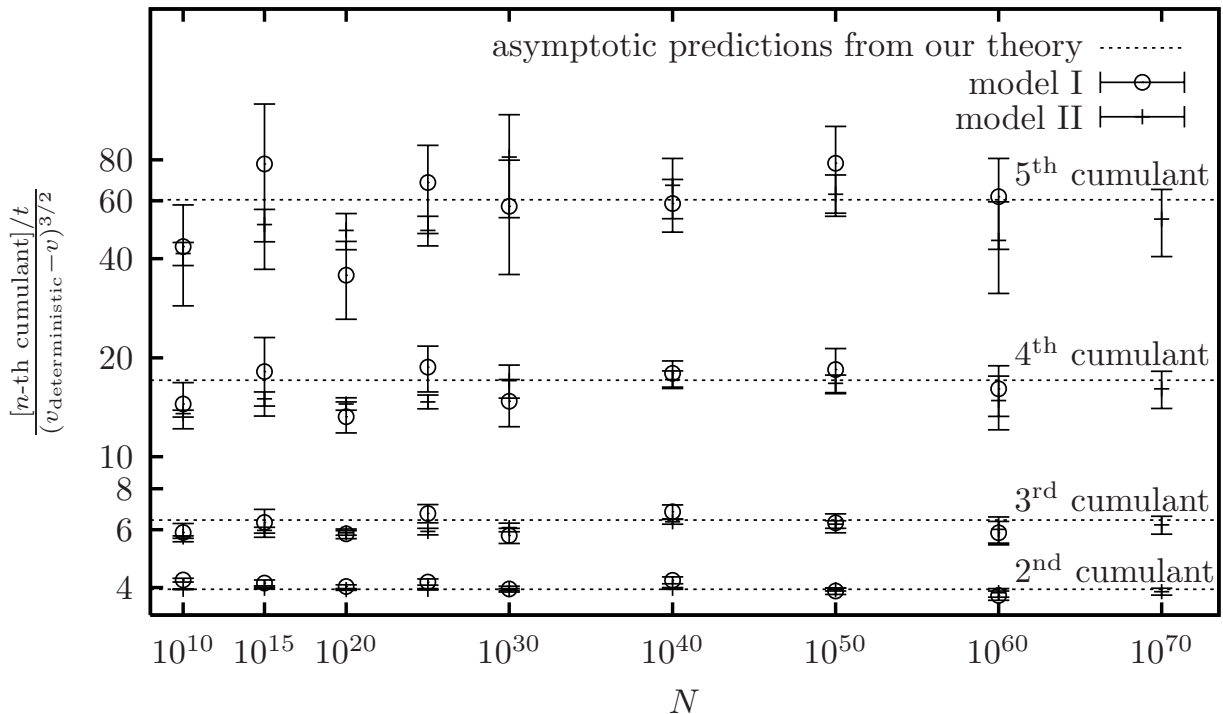}
\caption{\label{ratio}The ratio of cumulants 2 to 5 divided by the
correction to the velocity to the power $3/2$. The dashed lines are the
analytical prediction assuming only the cutoff theory (\ref{v_cutoff}) for
the velocity and the prediction (\ref{results}) for the cumulants.}
\end{figure*}

We define the position of the front at time $t$ by
\begin{equation}
X_t=\sum_{x=0}^\infty u(x,t).
\end{equation}
We start at time $t=0$ from the initial condition $u(x,0)=1$ for $x\leq 0$ and
$u(x,0)=0$ for $x>0$. We evolve it up to time $t=\log^2 N$ to get rid
of subasymptotic effects related to the building of the asymptotic shape of the 
front, and we measure the mean velocity between times 
$\log^2 N$ and $16\times \log^2 N$. 
For model I (many stochastic bins),  we average the results over $10^4$ such realizations.
For model II (only one stochastic bin), we generate  $10^5$ such realizations
for $N\leq 10^{50}$ and $10^4$ realizations for $N>10^{50}$.
In all our simulations, models~I and~II give
numerically indistinguishable results for the values of $N$ where both
models were simulated, as can be seen on the figures (results for model~I are
represented by a circle and for model~II by a cross).

\medbreak

First, we check that the effective width of the front
is $L_\text{eff}$ given by Eq.~(\ref{4th}).
We extract the latter from the measured
mean velocity $v$ using the formula
\begin{equation}
L_\text{eff}=\pi\sqrt{\frac{v^{\prime\prime}(\gamma_0)}{2\big(v(\gamma_0)-v\big)}}\ .
\label{Leff_from_v}
\end{equation}
We subtract from $L_\text{eff}$ 
the width of the front obtained in the cutoff theory 
$L=(\log N)/\gamma_0$,
and compare the numerical result with the analytical formula
\begin{equation}
L_\text{eff}-L
=\frac{3 \log(\log N)}{\gamma_0}+c+d \frac{\log(\log N)}{\log N}\ .
\label{ansatzLeff}
\end{equation}
The first term in the r.h.s. is suggested by our fourth assumption
(see Eq.~(\ref{4th})).
We have
added two subleading terms which go beyond our theory:
a constant term, and a term that vanishes at large $N$.
The latter are naturally expected to be the next terms in the
asymptotic expansion for large $N$. We include them
in this numerical analysis
because in the range of $N$ in which we are able to perform our numerical
simulations, they may still bring a significant contribution.

We fit (\ref{ansatzLeff})
to the numerical data obtained in the framework of
model~II, restricting ourselves to values of $N$ larger than $10^{30}$. 
In the fit, each data point is weighted by the statistical 
dispersion of its value in our sample of data.
We obtain a determination of the values of the free parameters
$c=-4.26\pm 0.01$ and $d=5.12\pm 0.27$, with a good quality of the fit 
($\chi^2/d.o.f\sim 1.15$).
The numerical data together with the theoretical predictions are 
shown in figure~\ref{loglogN}.
We see a clear convergence of the data to the predicted asymptotics
at large~$N$ (dotted line in the figure), 
but subleading corrections that we have accounted for phenomenologically
here are sizable over the whole range of $N$.

We now turn to the higher order cumulants. 
Our numerical data is shown in figure~\ref{cumulants} together with
the analytical
predictions obtained from~(\ref{results})
(dotted lines in the figure).
We see that the numerical simulations get very close to the analytical 
predictions at large $N$.
However, like in the case
of $L_\text{eff}$,
higher order corrections are presumably still important for
the lowest values of $N$ displayed on the plot.

We try to account for these corrections by replacing
the factor $(\log N)/\gamma_0=L$ in the denominator of the expression for the
cumulants in Eq.~(\ref{results})
by the Ansatz for $L_\text{eff}$ given in~(\ref{ansatzLeff}),
\emph{without retuning the parameters}.
The results are shown in figure~\ref{cumulants} (full lines), 
and are in excellent
agreement with the numerical data over the whole range of~$N$.
We could also
have refitted the parameters $c$ and $d$ for each cumulant separately, 
as, a priori, they are not predicted by our theory.
We observe that this is not required by our data.

This last observation suggests that all the cumulants can be computed,
with a good accuracy, with the effective width~$L_\text{eff}$ as the
only parameter. We check this on figure~\ref{ratio}, which
represents the ratio of the $n$-th cumulant (divided by time)
by the correction to the velocity $v_\text{deterministic}-v$ to
the power $3/2$. If one supposes that the correction to the velocity
varies like $1/L_\text{eff}^2$ and the cumulants like $1/L_\text{eff}^3$
for some effective width~$L_\text{eff}$, this width disappears from the
ratio plotted and one can compare the numerical results to our analytical
prediction with no free parameter or unknown subleading terms. Within
statistical error, the data seem to agree for $N$ large enough with
our prediction, suggesting that, indeed, all the cumulants can be
described with a good accuracy with only the effective
width~$L_\text{eff}$.

Simulations, not shown here, for the model introduced in~\cite{BD2}
support also our predictions~(\ref{results}).

\section{Conclusion}%

The main idea that we have put forward in the present work is that all
the fluctuations of the front position, and in particular the diffusion
constant, are dominated by large but rare fluctuations at the tip of the
front.

Under some more precise assumptions (hypotheses of section~\ref{sec:deriv})
on these fluctuations, we were able to obtain explicit expressions
(\ref{results}) of the cumulants of the position of the front.
We checked these predictions in our numerical simulations of
section~\ref{sec:num}. In section~\ref{sec:justification}, we gave some
arguments in support of the four hypotheses of section~\ref{sec:deriv}.
None of these arguments can be regarded as a mathematical derivation, and
we can imagine that some details, such as the precise shape of the
distribution of fluctuations (\ref{distlen}) or the explicit expression
(\ref{Rd}), could be slightly modified by a more precise
analysis. We believe however, given the good agreement of the
predictions (\ref{results}) with the numerical simulations, that our
picture is very close, if not identical, to the actual behavior of the
front for large values of $N$.

To conclude, we would like to point out the remarkable similarity between
the predictions~(\ref{results})
and the exact results obtained
recently\cite{BrunetDerrida.04} in the context of directed polymers.
Basically, the results of~\cite{BrunetDerrida.04} are the
same, \textit{mutatis mutandis} as our present results~(\ref{results}),
for all the cumulants. The only significant change is that the $3\log\log
N$ for the velocity and the $1/\log^3 N$ dependence for all the cumulants
in~(\ref{results}) corresponds, in \cite{BrunetDerrida.04}, to a
$\log\log N$ for the velocity and a $1/\log N$ for all the
cumulants\cite[Eq.~(23) with $L=\log N$. The term $L+\log L$ in the
velocity corresponds to $v_\text{cutoff}$, see
Eq.~(28)]{BrunetDerrida.04}. What is interesting is that our scenario
of section~\ref{sec:deriv} for FKPP fronts applies also for
the system studied in~\cite{BrunetDerrida.04}: Indeed, the
fluctuations of the position are mainly due to the rare big events taking
place at the tip of the ``front'', \cite[last paragraph before
conclusion]{BrunetDerrida.04}, the position of the rightmost particle is
given by (\ref{distlen}) \cite[Eq.~(32) with $\delta=-\log q$ and
$X_t=\log B_t$]{BrunetDerrida.04}, the effect of a large fluctuation can
be written as (\ref{Rd}) with the $L^3$ term replaced by $L$ \cite[the
log of~(34) can be written as $X_{t+1}-X_t=L +\log L +
R(\delta)$]{BrunetDerrida.04}, relevant fluctuation (of size $\log L$
instead of $3\log L$) appears every $L$ timesteps (instead of every $L^3$
timesteps) and the relaxation time is 1 instead of $L^2$.
This similarity
may add a
further piece of evidence for our results.

\bigbreak

This work was partially supported by the US Department of Energy.

\appendix

\section{Limit~$N\to\infty$}\label{infinite}%

In this appendix, we try to provide an argument for the exponential
decay~(\ref{distlen}) of the distribution for the width of the front.
To this aim, we consider a very simple model of reaction-diffusion:
particles diffuse on the line and during each time interval~$dt$, each
particle duplicates with a probability~$dt$. The motions of all
the particles are uncorrelated.

If one added a saturation rule as described at the beginning of
section~\ref{sec:deriv}, the density of particles (or the number of
particles on the right of $x$, depending on the precise saturation rule)
would be described by a stochastic
FKPP equation.
However, the saturation affects only the motion of particles
in the bulk of the front, where the density is high.
As the fluctuations develop in the low density region, it is reasonable
to assume that the distribution of the size of the fluctuations are well
described by the model \emph{without any saturation}.

For this model without saturation,
let~$P_t(x)$ the probability that, at time~$t$, no particles are present
on the right of~$x$ given that, at $t=0$, there is a single particle at
the origin: $P_0(x)=\theta(x)$. During the first ``time step'' $dt$, the
only particle in the system moves by a quantity~$\eta\sqrt{dt}$
where~$\eta$ is a Gaussian number of variance~$2$, and duplicates with a
probability~$dt$. If it duplicates, the probability~$P_{t+dt}(x)$ is the
probability that the offsprings of both particles are on the left of~$x$.
As the particles have uncorrelated motion, this is the product of the
probabilities for each offspring. Finally, one gets\cite{McKean.75}
\begin{equation*}
P_{t+dt}(x) = \left\langle P_t(x-\eta\sqrt{dt})(1-dt) 
	+  P_t^2(x-\eta\sqrt{dt}) dt\right\rangle,
\end{equation*}
where the average is on~$\eta$. After simplification,
\begin{equation}
\partial_t P = \partial_x^2 P -P+P^2.
\end{equation}
One notices that $1-P_t(x)$ is solution of the deterministic FKPP
equation~(\ref{FKPP}). Therefore, for large~$t$ and $x$,
\cite{Bramson.83,BrunetDerrida.97,VanSaarloos.03}
\begin{equation*}
1-P_t(x) \sim z e^{-z-\frac{z^2}{4t}}\text{\quad for }z=x-2t
+\frac{3}{2}\log t.
\end{equation*}
Let~$Q_t(x)$ be the probability that there are no particles on the right
of~$x$ when the initial condition is a given density of
particles~$\rho_0(x)$. Using the fact that all the particles are
independent, one gets easily
\begin{equation}
Q_t(x)=\exp\left[-\int dy\ \rho_0(y)\Big(1-P_t(x-y)\Big)\right].
\label{QP}
\end{equation}
$\rho_0(y)$ needs to
reproduce the shape of the front seen from the tip.
Starting from (\ref{h_cutoff}), we write~$\rho(y)=N h_\text{cutoff}(L+y)$
and take the large~$N$ limit. One gets $\rho_0(y)=-y\exp(-y)$ for~$y<0$
and $\rho_0(y)=0$ for $y>0$. Evaluating the integral in (\ref{QP}), one
gets, for large~$t$ and $x-2t\ll\sqrt t$,
\begin{equation}
Q_t(x)\approx\exp\left[-C e^{-(x-2t)}\right].
\label{gumbel}
\end{equation}
(Notice how the $(3/2)\log t$ factor canceled out). The probability
distribution function of the rightmost particle is clearly~$\partial_x
Q_t(x)$. We see that in this stochastic model, the front moves at a
deterministic velocity equal to~$2$ and that the position of the
rightmost particle around the position of the front is given by a Gumbel
distribution.

The distribution~(\ref{gumbel}) gives our first hypothesis~(\ref{distlen})
for large fluctuations ($\delta=x-2t\gg1$). Our attempts to check
numerically (\ref{gumbel}) by simulating fronts with a large but finite
number of particles confirmed this exponential decay for
large~$\delta$, but showed some discrepancy for $\delta<0$, which we do
not understand. This, however, does not affect the
hypothesis~(\ref{distlen}).

\section{Moving wall}\label{wall}%

We consider again the reaction-diffusion model introduced in
appendix~\ref{infinite}. As we said, one needs to add a saturation effect
to obtain a propagating front equation for the density, but doing so introduces
correlations in the motions of the particles
that make the model hard to solve.
In this appendix, we introduce an approximate way of adding a
saturation effect which does not introduce any such correlation.

In a real front, the tip is subject to huge fluctuations happening on
short time scales. On the other hand, the bulk of the front moves
smoothly and adjusts very slowly to the fluctuations happening at the tip.
Therefore, we believe that, for times not too large, it is a reasonable
approximation to assume that the bulk of the front moves at a constant
velocity. 

To implement this idea, our model is the following: a wall starting at
the origin is moving to the right at a constant velocity~$v$.
Particles are present on the right of the wall. The particles are
evolving as in appendix~\ref{infinite}, except that whenever a particle
crosses the wall, it is removed. 

We first consider a single particle starting at a distance~$z$ of the
wall.
After a time $t$, either all the offsprings of this particle have been
caught by the wall, or some have survived.
We want to compute the probability~$E_t(z)$ that all the particles
have been caught at time~$t$. The original particle, after a time~$dt$, is
at a distance~$z-vdt+\eta\sqrt{dt}$ from the wall, and it might have
duplicated with probability~$dt$. Using the same method as in
appendix~\ref{infinite}, one gets
\begin{equation}
\partial_t E_t=\partial_z^2 E_t -v\partial_z E_t -E_t+E_t^2.
\end{equation}
with the conditions
\begin{equation}
E_0(z)=0\text{ for $z>0$}\quad\text{and}\quad E_t(z)=1\text{ for $z<0$}.
\end{equation}
In the long time limit, $E_t(z)$ converges to the stationary solution
$E_\infty(z)$, and one
recognizes that~$h(z)=1-E_\infty(-z)$ is
the stationary solution of the FKPP equation~(\ref{FKPP}) if~$z=x-vt$.
In other words, $1-E_\infty(-z)$ is the shape of a travelling front. As this
shape reaches~$0$ for~$z=0$, it must be a front with a sine arch and a
velocity~$v$ smaller than~$2$, as in (\ref{cutoff}). So
if~$v<2$, the probability~$1-E_\infty(-z)$ is the shape of the
front with a cutoff:
\begin{equation}
1-E(z)\sim L e^{-L}\sin\left(\pi \frac{z}{L}\right)e^{z}
\quad\text{where}\quad v=2-\frac{\pi^2}{L^2}.
\label{1-E}
\end{equation}
(The extra factor $e^{-L}$ comes from the fact that~$z=0$ is the tip of
the front in (\ref{1-E}) while it is the bulk of the front in
(\ref{h_cutoff}). If $v>2$, all the particles eventually
die and $E_t(z)$ converges to~$1$.)

If one starts with a density~$\rho(z)$ of particles at time~$t=0$, the
probability~$E^*_t$ that everybody dies is given, similarly to~(\ref{QP}), by
\begin{equation}
E^*_t=\exp\left[-\int_{0}^{+\infty} dz\ \rho(z)\Big(1-E_t(z)\Big)\right].
\end{equation}
We consider, as an initial condition, the situation in the real front
with~$\rho(z)=N h(z)\sim N L \sin(\pi z/L)\exp(-z)$ for~$z<L$
as in (\ref{h_cutoff}). One gets, for long
times,
\begin{equation}
E^*_t\to \exp\left[-{C} N L^3 e^{-L}\right].
\end{equation}
We see that the system survives if
\begin{equation}
N L^3 e^{-L}\gtrapprox 1\quad\text{or}\quad L\lessapprox \log N +3\log\log N,
\end{equation}
which, given~$\gamma_0=1$, is exactly~(\ref{4th}).


\begin{thebibliography}{31}
\expandafter\ifx\csname natexlab\endcsname\relax\def\natexlab#1{#1}\fi
\expandafter\ifx\csname bibnamefont\endcsname\relax
  \def\bibnamefont#1{#1}\fi
\expandafter\ifx\csname bibfnamefont\endcsname\relax
  \def\bibfnamefont#1{#1}\fi
\expandafter\ifx\csname citenamefont\endcsname\relax
  \def\citenamefont#1{#1}\fi
\expandafter\ifx\csname url\endcsname\relax
  \def\url#1{\texttt{#1}}\fi
\expandafter\ifx\csname urlprefix\endcsname\relax\def\urlprefix{URL }\fi
\providecommand{\bibinfo}[2]{#2}
\providecommand{\eprint}[2][]{\url{#2}}

\bibitem[{\citenamefont{Fisher}(1937)}]{FKPP}
\bibinfo{author}{\bibfnamefont{R.~A.} \bibnamefont{Fisher}},
  \bibinfo{journal}{Annals of Eugenics} \textbf{\bibinfo{volume}{7}},
  \bibinfo{pages}{355} (\bibinfo{year}{1937});
\newblock
\bibinfo{author}{\bibfnamefont{A.}~\bibnamefont{Kolmogorov}},
  \bibinfo{author}{\bibfnamefont{I.}~\bibnamefont{Petrovsky}},
  \bibnamefont{and}
  \bibinfo{author}{\bibfnamefont{N.}~\bibnamefont{Piscounov}},
  \bibinfo{journal}{Bull. Univ. \'Etat Moscou, A} \textbf{\bibinfo{volume}{1}},
  \bibinfo{pages}{1} (\bibinfo{year}{1937}).

\bibitem[{\citenamefont{van Saarloos}(2003)}]{VanSaarloos.03}
\bibinfo{author}{\bibfnamefont{W.}~\bibnamefont{van Saarloos}},
  \bibinfo{journal}{Phys. Rep.} \textbf{\bibinfo{volume}{386}},
  \bibinfo{pages}{29} (\bibinfo{year}{2003}).

\bibitem[{\citenamefont{Pechenik and Levine}(1999)}]{PechenikLevine.99}
\bibinfo{author}{\bibfnamefont{L.}~\bibnamefont{Pechenik}} \bibnamefont{and}
  \bibinfo{author}{\bibfnamefont{H.}~\bibnamefont{Levine}},
  \bibinfo{journal}{Phys. Rev. E} \textbf{\bibinfo{volume}{59}},
  \bibinfo{pages}{3893} (\bibinfo{year}{1999}).

\bibitem[{\citenamefont{Doering et~al.}(2003)\citenamefont{Doering, Mueller,
  and Smereka}}]{DoeringMuellerSmereka.03}
\bibinfo{author}{\bibfnamefont{C.~R.} \bibnamefont{Doering}},
  \bibinfo{author}{\bibfnamefont{C.}~\bibnamefont{Mueller}}, \bibnamefont{and}
  \bibinfo{author}{\bibfnamefont{P.}~\bibnamefont{Smereka}},
  \bibinfo{journal}{Physica A} \textbf{\bibinfo{volume}{325}},
  \bibinfo{pages}{243} (\bibinfo{year}{2003}).

\bibitem[{\citenamefont{Majumdar and Krapivsky}(2003)}]{MajumdarKrapivsky.03}
\bibinfo{author}{\bibfnamefont{S.~N.} \bibnamefont{Majumdar}} \bibnamefont{and}
  \bibinfo{author}{\bibfnamefont{P.~L.} \bibnamefont{Krapivsky}},
  \bibinfo{journal}{Phys. A} \textbf{\bibinfo{volume}{318}},
  \bibinfo{pages}{161} (\bibinfo{year}{2003}).

\bibitem[{\citenamefont{Derrida and Spohn}(1988)}]{DerridaSpohn.Polymers.88}
\bibinfo{author}{\bibfnamefont{B.}~\bibnamefont{Derrida}} \bibnamefont{and}
  \bibinfo{author}{\bibfnamefont{H.}~\bibnamefont{Spohn}}, \bibinfo{journal}{J.
  Stat. Phys.} \textbf{\bibinfo{volume}{51}}, \bibinfo{pages}{817}
  (\bibinfo{year}{1988}).

\bibitem[{\citenamefont{Carpentier and {Le
  Doussal}}(2000)}]{CarpentierLedoussal.00}
\bibinfo{author}{\bibfnamefont{D.}~\bibnamefont{Carpentier}} \bibnamefont{and}
  \bibinfo{author}{\bibfnamefont{P.}~\bibnamefont{{Le Doussal}}},
  \bibinfo{journal}{Nucl. Phys. B} \textbf{\bibinfo{volume}{588}},
  \bibinfo{pages}{531} (\bibinfo{year}{2000}).

\bibitem[{\citenamefont{Munier and Peschanski}(2003)}]{MunierPeschanski.03}
\bibinfo{author}{\bibfnamefont{S.}~\bibnamefont{Munier}} \bibnamefont{and}
  \bibinfo{author}{\bibfnamefont{R.}~\bibnamefont{Peschanski}},
  \bibinfo{journal}{Phys. Rev. Lett.} \textbf{\bibinfo{volume}{91}},
  \bibinfo{pages}{232001} (\bibinfo{year}{2003}).

\bibitem[{\citenamefont{Balitsky}(1996)}]{BK}
\bibinfo{author}{\bibfnamefont{I.}~\bibnamefont{Balitsky}},
  \bibinfo{journal}{Nucl. Phys. B} \textbf{\bibinfo{volume}{463}},
  \bibinfo{pages}{99} (\bibinfo{year}{1996});
\newblock
\bibinfo{author}{\bibfnamefont{Y.~V.} \bibnamefont{Kovchegov}},
  \bibinfo{journal}{Phys. Rev. D} \textbf{\bibinfo{volume}{60}},
  \bibinfo{pages}{034008} (\bibinfo{year}{1999});
\newblock
\bibinfo{journal}{Phys. Rev. D} \textbf{\bibinfo{volume}{61}},
  \bibinfo{pages}{074018} (\bibinfo{year}{2000}).

\bibitem[{\citenamefont{Marquet et~al.}(2005)\citenamefont{Marquet, Peschanski,
  and Soyez}}]{Marquet.05}
\bibinfo{author}{\bibfnamefont{C.}~\bibnamefont{Marquet}},
  \bibinfo{author}{\bibfnamefont{R.}~\bibnamefont{Peschanski}},
  \bibnamefont{and} \bibinfo{author}{\bibfnamefont{G.}~\bibnamefont{Soyez}},
  \bibinfo{journal}{Nucl. Phys. A} \textbf{\bibinfo{volume}{756}},
  \bibinfo{pages}{399} (\bibinfo{year}{2005}).

\bibitem[{\citenamefont{Aronson and Weinberger}(1975)}]{AronsonWeinberger.75}
\bibinfo{author}{\bibfnamefont{D.~G.} \bibnamefont{Aronson}} \bibnamefont{and}
  \bibinfo{author}{\bibfnamefont{H.~F.} \bibnamefont{Weinberger}},
  \bibinfo{journal}{Lecture Notes in Mathematics}
  \textbf{\bibinfo{volume}{446}}, \bibinfo{pages}{5} (\bibinfo{year}{1975}).

\bibitem[{\citenamefont{Brunet and Derrida}(1997)}]{BrunetDerrida.97}
\bibinfo{author}{\bibfnamefont{{\'E}.}~\bibnamefont{Brunet}} \bibnamefont{and}
  \bibinfo{author}{\bibfnamefont{B.}~\bibnamefont{Derrida}},
  \bibinfo{journal}{Phys. Rev. E} \textbf{\bibinfo{volume}{56}},
  \bibinfo{pages}{2597} (\bibinfo{year}{1997}).

\bibitem[{\citenamefont{Enberg et~al.}(2005)\citenamefont{Enberg,
  Golec-Biernat, and Munier}}]{Enberg.05}
\bibinfo{author}{\bibfnamefont{R.}~\bibnamefont{Enberg}},
  \bibinfo{author}{\bibfnamefont{K.}~\bibnamefont{Golec-Biernat}},
  \bibnamefont{and} \bibinfo{author}{\bibfnamefont{S.}~\bibnamefont{Munier}},
  \bibinfo{journal}{Phys. Rev. D} \textbf{\bibinfo{volume}{72}},
  \bibinfo{pages}{074021} (\bibinfo{year}{2005}).

\bibitem[{\citenamefont{Kessler et~al.}(1998)\citenamefont{Kessler, Ner, and
  Sander}}]{Kessler.98}
\bibinfo{author}{\bibfnamefont{D.~A.} \bibnamefont{Kessler}},
  \bibinfo{author}{\bibfnamefont{Z.}~\bibnamefont{Ner}}, \bibnamefont{and}
  \bibinfo{author}{\bibfnamefont{L.~M.} \bibnamefont{Sander}},
  \bibinfo{journal}{Phys. Rev. E} \textbf{\bibinfo{volume}{58}},
  \bibinfo{pages}{107} (\bibinfo{year}{1998}).

\bibitem[{\citenamefont{Panja}(2003)}]{Panja}
\bibinfo{author}{\bibfnamefont{D.}~\bibnamefont{Panja}},
  \bibinfo{journal}{Phys. Rev. E} \textbf{\bibinfo{volume}{68}},
  \bibinfo{pages}{065202(R)} (\bibinfo{year}{2003});
\newblock
\bibinfo{journal}{Phys. Rep.} \textbf{\bibinfo{volume}{393}},
  \bibinfo{pages}{87} (\bibinfo{year}{2004}).

\bibitem[{\citenamefont{Mueller and Sowers}(1995)}]{MuellerSowers.95}
\bibinfo{author}{\bibfnamefont{C.}~\bibnamefont{Mueller}} \bibnamefont{and}
  \bibinfo{author}{\bibfnamefont{R.~B.} \bibnamefont{Sowers}},
  \bibinfo{journal}{J. Funct. Anal.} \textbf{\bibinfo{volume}{128}},
  \bibinfo{pages}{439} (\bibinfo{year}{1995}).

\bibitem[{\citenamefont{Brunet and Derrida}(1999)}]{BD2}
\bibinfo{author}{\bibfnamefont{{\'E}.}~\bibnamefont{Brunet}} \bibnamefont{and}
  \bibinfo{author}{\bibfnamefont{B.}~\bibnamefont{Derrida}},
  \bibinfo{journal}{Computer Physics Communications}
  \textbf{\bibinfo{volume}{121--122}}, \bibinfo{pages}{376}
  (\bibinfo{year}{1999});
\newblock
\bibinfo{journal}{J. Stat. Phys.} \textbf{\bibinfo{volume}{103}},
  \bibinfo{pages}{269} (\bibinfo{year}{2001}).

\bibitem[{\citenamefont{Mueller and Shoshi}(2004)}]{MuellerShoshi.04}
\bibinfo{author}{\bibfnamefont{A.~H.} \bibnamefont{Mueller}} \bibnamefont{and}
  \bibinfo{author}{\bibfnamefont{A.~I.} \bibnamefont{Shoshi}},
  \bibinfo{journal}{Nucl. Phys. B} \textbf{\bibinfo{volume}{692}},
  \bibinfo{pages}{175} (\bibinfo{year}{2004}).

\bibitem[{\citenamefont{Colon and Doering}(2005)}]{ColonDoering.05}
\bibinfo{author}{\bibfnamefont{J.~G.} \bibnamefont{Colon}} \bibnamefont{and}
  \bibinfo{author}{\bibfnamefont{C.~R.} \bibnamefont{Doering}},
  \bibinfo{journal}{J. Stat. Phys.} \textbf{\bibinfo{volume}{120}},
  \bibinfo{pages}{421} (\bibinfo{year}{2005}).

\bibitem[{\citenamefont{Iancu et~al.}(2005)\citenamefont{Iancu, Mueller, and
  Munier}}]{IancuMuellerMunier.05}
\bibinfo{author}{\bibfnamefont{E.}~\bibnamefont{Iancu}},
  \bibinfo{author}{\bibfnamefont{A.~H.} \bibnamefont{Mueller}},
  \bibnamefont{and} \bibinfo{author}{\bibfnamefont{S.}~\bibnamefont{Munier}},
  \bibinfo{journal}{Phys. Lett. B} \textbf{\bibinfo{volume}{606}},
  \bibinfo{pages}{342} (\bibinfo{year}{2005}).

\bibitem[{\citenamefont{Moro}()}]{Moro_private}
\bibinfo{author}{\bibfnamefont{E.}~\bibnamefont{Moro}}, \bibinfo{note}{private
  discussion}.

\bibitem[{\citenamefont{Bramson}(1983)}]{Bramson.83}
\bibinfo{author}{\bibfnamefont{M.~D.} \bibnamefont{Bramson}},
  \bibinfo{journal}{Mem. Am. Math. Soc.} \textbf{\bibinfo{volume}{44}}
  (\bibinfo{year}{1983}).

\bibitem[{\citenamefont{Moro}(2004{\natexlab{a}})}]{Moro}
\bibinfo{author}{\bibfnamefont{E.}~\bibnamefont{Moro}}, \bibinfo{journal}{Phys.
  Rev. E} \textbf{\bibinfo{volume}{69}}, \bibinfo{pages}{060101(R)}
  (\bibinfo{year}{2004}{\natexlab{a}});
\newblock
\bibinfo{journal}{Phys.
  Rev. E} \textbf{\bibinfo{volume}{70}}, \bibinfo{pages}{045102(R)}
  (\bibinfo{year}{2004}{\natexlab{b}}).

\bibitem[{\citenamefont{Brunet and Derrida}(2004)}]{BrunetDerrida.04}
\bibinfo{author}{\bibfnamefont{{\'E}.}~\bibnamefont{Brunet}} \bibnamefont{and}
  \bibinfo{author}{\bibfnamefont{B.}~\bibnamefont{Derrida}},
  \bibinfo{journal}{Phys. Rev. E} \textbf{\bibinfo{volume}{70}},
  \bibinfo{pages}{016106} (\bibinfo{year}{2004}).

\bibitem[{\citenamefont{McKean}(1975)}]{McKean.75}
\bibinfo{author}{\bibfnamefont{H.~P.} \bibnamefont{McKean}},
  \bibinfo{journal}{Comm. Pure Appl. Math.} \textbf{\bibinfo{volume}{28}},
  \bibinfo{pages}{323} (\bibinfo{year}{1975}).

\end{thebibliography}
\end{document}